\documentclass[aps,prapplied,reprint,floatfix,longbibliography,superscriptaddress,amsmath,amssymb,compress]{revtex4-2}
\pdfoutput=1

\usepackage{graphicx}
\usepackage{color}
\usepackage{hyperref}

\usepackage{booktabs}

\usepackage{siunitx}

\AtBeginDocument{
  \heavyrulewidth=.08em
  \lightrulewidth=.06em
  \cmidrulewidth=.03em
  \belowrulesep=.65ex
  \belowbottomsep=0pt
  \aboverulesep=.4ex
  \abovetopsep=0pt
  \cmidrulesep=\doublerulesep
  \cmidrulekern=.5em
  \defaultaddspace=.5em
  }

\newcommand{\papertitle}{Full-band reciprocal parametric amplification in coupled oscillator arrays}

\newcommand{\acktext}{Research reported in this
  publication was supported by the National Science Foundation through award DMR-2145766 and award CMMI-2128671.
  We thank Benjam\'in Alem\'an and Uriel Hernandez for insights on the experimental system.
  JP acknowledges insightful discussions with Arnaud Lazarus.}

\newcommand{\wmin}{\omega_\mathrm{min}}
\newcommand{\wmax}{\omega_\mathrm{max}}
\newcommand{\wmid}{\tilde{\omega}}
\newcommand{\staticK}{\mathbf{K}_0}
\newcommand{\modK}{\mathbf{K}_1}
\newcommand{\basisvec}{\hat{e}}

\newcommand{\omegae}{\omega_{\mathrm{e}}}

\newcommand{\eqnref}[1]{Eq.~\eqref{#1}}
\newcommand{\Equationref}[1]{Equation~\eqref{#1}}
\newcommand{\appref}[1]{Appendix~\ref{#1}}
\newcommand{\subfiglabel}[1]{{(#1)}}
\newcommand{\figref}[1]{Fig.~\ref{#1}}
\newcommand{\subfigref}[2]{\figref{#1}\subfiglabel{#2}}
\newcommand{\secref}[1]{Sec.~\ref{#1}}

\begin{document}
\title{\papertitle}
\author{Benjamin Kauffman}
\affiliation{Institute for Fundamental Science, Materials Science Institute, and Department of Physics,
University of Oregon, Eugene, OR 97403}
\author{Jayson Paulose}
\email{jpaulose@uoregon.edu}
\affiliation{Institute for Fundamental Science, Materials Science Institute, and Department of Physics,
University of Oregon, Eugene, OR 97403}

\begin{abstract}

Parametric resonance provides a versatile mechanism to manipulate wave signals in time-modulated metamaterials, but typically operates within narrow wavelength or frequency ranges.
Here, we show that one-dimensional arrays of coupled resonators can generically exhibit parametric amplification across an entire band under standing-wave parametric modulation.
We develop a general physical framework to describe the parametric resonance of isolated bands, derive modulation conditions needed to achieve full-band amplification for a given band, and demonstrate their validity in a model of active mechanical metamaterials based on tunable membrane resonator arrays.
We numerically demonstrate applications of the proposed strategy in distortion-free amplification and temporal splitting of wave packets, with results in quantitative agreement with our theoretical calculations. 
Our strategies for broadband amplification can be generalized to other active wave media with nearly flat or cosine-like dispersion relations, with potential applications in loss-mitigation, noise squeezing, and programmable wave manipulation.

\end{abstract}

\maketitle

\section{Introduction}

\begin{figure*}
  \centering
  \includegraphics[width=\textwidth]{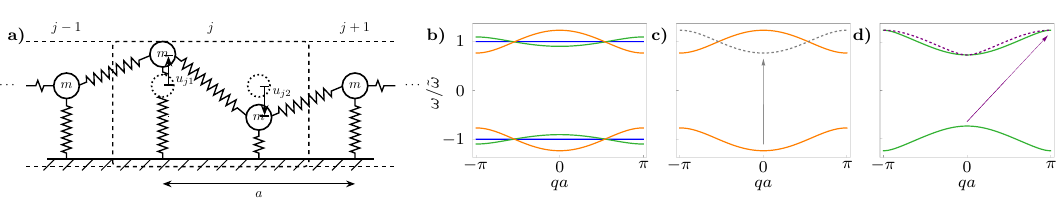}
  \caption{
    \subfiglabel{a} Abstract model: An array of oscillators with arbitrary couplings between them, periodic in space with unit cells indexed by $j$. A vector $\mathbf{u}=(u_{j1},u_{j2},\ldots,u_{jn})^T$ encodes the displacements of masses in unit cell $j$. Here, $n=2$ and the displacements of the masses are labeled as $u_{j1}$ and $u_{j2}$. 
    \subfiglabel{b} Characteristic cosine-like frequency-wavevector dispersion relation of the lowest band $\omega(q)$ and its partner $-\omega(q)$ with three different characteristic band shapes. Orange: Optical-like band with negative slope at positive wavevector $q$. Blue: Flat band with constant frequency across $q$. Green: Acoustic-like band with positive slope at positive $q$.
    \subfiglabel{c} Modulation scheme that generates parametric resonances across a nearly flat band with average frequency $\wmid$.
    Dashed curve denotes Floquet-Bloch replica of the lower branch generated by a uniform modulation, obtained by copying and displacing the band vertically by the modulation frequency $\Omega$ (arrow).
    Here, the choice $\Omega = 2\wmid$ brings the replica close to the positive branch across all wavevectors, provided the bandwidth is low.
    \subfiglabel{d} Modulation with $\pi$ phase offset between neighboring unit cells displaces the replica (dashed curve) in both frequency and wavevector directions (purple arrow), bringing the band to nearly coincide with its replica at all $q$ despite the large bandwidth.
  }
  \label{fig:intro}
\end{figure*}

The periodic time-modulation of local parameters in oscillatory systems provides mechanisms for transferring energy into and across normal modes~\cite{louisell} that complement the direct forcing of these modes.
Such \emph{parametric phenomena} provide a basis for designing active waveguides for sound and electromagnetic signals with capabilities that cannot be achieved in passive structures~\cite{Zangeneh-Nejad2019,Wang2020,Yin2022}, such as nonreciprocal wave transmission~\cite{Sounas2017,Nassar2020} and amplification~\cite{Koutserimpas2018a,Trainiti2019,Aumentado2020,Kim2023}.
However, parametric phenomena typically operate in narrow frequency and/or wavevector bands set by the modulation frequency due to strict resonance conditions, which limits their utility for processing signals with rich spectra such as multiplexed signals and wave packets.
Mechanisms that allow the parametric resonance conditions to be satisfied across entire frequency bands could therefore expand the potential applications of parametric modulation for signal manipulation.
 
In systems with a linear frequency-wavevector (dispersion) relation, broadband amplification can be generated by setting up a traveling-wave modulation with speed equal to the constant wave velocity (the so-called sonic or luminal limit~\cite{Cassedy1963,Cassedy1967,Galiffi2019,Kruss2022}).
However, this approach excludes the vast multitude of coupled-oscillator systems that exhibit zero-wavevector bandgaps due to a characteristic resonance frequency of individual oscillators.
Furthermore, traveling-wave parametric amplification is generically nonreciprocal as the propagating modulation wave breaks directional symmetry in the structure: waves propagating in one direction are boosted while the other direction is unaffected.
While nonreciprocity can be desirable in some contexts, isotropic full-band amplification is a useful functionality for processing signals with complex spectral content irrespective of propagation direction.

Here, we achieve \emph{reciprocal} amplification of entire \emph{gapped} bands in active metamaterials based on coupled resonator arrays.
Our approach exploits the fact that 1D resonator arrays generically exhibit cosine-like dispersion relations due to nearest-neighbor couplings between identical resonator elements, whether the degrees of freedom are mechanical (Ref.~\onlinecite{Karki2021} and \figref{fig:intro}), optical (coupled-resonator optical waveguides for light~\cite{Yariv1999}) or plasma modes (Josephson junction arrays for microwaves~\cite{Hutter2011}).
We investigate two complementary approaches: a spatially-uniform periodic parametric modulation, which generates fully-amplified flat bands (\subfigref{fig:intro}{c}); and a standing-wave modulation, which generates amplified bands with a cosine-like dispersion relation that preserves the static bandwidth (\subfigref{fig:intro}{d}).
In both cases, we identify the relevant features of the static dispersion relation and the modulation field that influence the full-band amplification, and provide analytical predictions for the strength of the effect from these features.
We also connect the full-band amplification to wave packet dynamics, illustrating tunable amplification and splitting of wave packets as potential applications of the fully amplified bands.

Our approach to generating full-band reciprocal parametric amplification is schematically outlined in \subfigref{fig:intro}{c--d} for uniform and standing-wave modulations respectively.
We consider a periodic array of harmonic oscillators (such as the fundamental vibrational modes of membrane resonators) with spacing $a$. Decomposing the collective vibrations of the array into wavelike modes $\propto e^{i(qx-\omega t)}$ generically yields $n$ modes per wavevector $q$, where $n$ is the number of oscillators (and thus the number of degrees of freedom) in a repeating unit cell of the array. 
Nearest-neighbor couplings broaden these modes into $n$ bands $\omega_\beta(q)$ defined on a Brillouin zone $-\pi/a < q \leq \pi/a$ of unique wavevectors $q$, each with zero slope at the band center and edges, which makes the resulting bands cosine-like in shape. Since the oscillations are governed by dynamical equations that are second-order in time, the complete mode space is described by pairs of bands $\pm \omega_\beta(q)$ for each $\beta$. 
We focus on a single band $\omega_\beta(q)$ that is gapped from other bands, and seek to generate parametric resonances across the entire band by applying time-periodic modulations of oscillator frequencies and/or coupling parameters.

When the modulation is uniform across the array, the existence of parametric resonances can be predicted by making copies (so-called Floquet-Bloch replicas) of the static bands displaced vertically by the modulation frequency $\Omega$, as shown schematically in \subfigref{fig:intro}{c}. Overlaps of the positive branch of the band with a direct copy of its negative branch indicate a degenerate parametric resonance where the modulation frequency is exactly twice the band frequency~\cite{YakubovichStarzhinskii,Cassedy1963}. In the $\delta \to 0$ limit, parametric resonance only occurs if the copy exactly coincides with the original band, which occurs when the band is completely flat and the modulation frequency is set to twice the flat band frequency.
Bands that deviate from perfect flatness with limited bandwidth can also be parametrically amplified by tuning the modulation frequency to bring the Floquet-Bloch replica as close as possible to the original band, provided the modulation strength is large enough to overcome the mismatch between the band and its copy at all wavevectors.
However, the frequency of the oscillatory component of all the resulting amplified modes is restricted to $\Omega/2$; i.e. the amplified band has an exactly flat real dispersion relation.

To induce parametric resonances that maintain the frequency range and cosine-like dispersion relation of the static band, we use a standing-wave modulation with spatial wavelength set to twice the array spacing. 
The resulting Floquet-Bloch replicas are displaced by the modulation wavevector $\pi/a$ in the quasimomentum direction~\cite{Cassedy1963,Galiffi2019}; we can then choose $\Omega$ to induce a near-overlap of the replicas of the lower and upper branches, as shown in \subfigref{fig:intro}{d}.
Generically, the overlap will not be perfect because the band is not necessarily symmetric under the resulting transformation,
but we expect that the entire band can still be brought into resonance with its partner at sufficiently large modulation amplitudes to overcome small mismatches between the band and its copy.
We also note that the resonance preserves $q \to -q$ symmetry on the periodic wavevector space, ensuring that the effect does not break reciprocity of wave transport in the system.

In the remainder of this paper, we will translate these qualitative observations into a quantitative theory of full-band parametric amplification.
We derive resonance conditions---modulation strengths needed to overcome the deviations from perfect overlap in the Floquet-Bloch replicas---that can be expressed as a single Mathieu equation or a pair of coupled Mathieu equations, with parameters determined by the Bloch eigenstates of the static bands.
The reduction to Mathieu oscillators allows us to derive expressions for quantities of interest using well-established techniques~\cite{mathieubook,nayfehmook,Kovacic2018}.
While our framework can be generally applied to coupled parametric oscillator arrays in one dimension,
we demonstrate the applicability of our results using a discrete mechanical model of membrane-resonator based phononic metamaterials~\cite{Karki2021,Karki2023} (\subfigref{fig:intro}{a}).
We also illustrate potential applications of both flat and finite-bandwidth amplified bands for wave packet manipulation using dynamical simulations: we show that propagating wave packets can be amplified with minimal distortion, and can be split into counterpropagating pairs by parametric pulses (temporal wave packet splitting).
The designed features of the amplified bands quantitatively predict the simulated wave packet dynamics.
Since the framework depends only on the static band structure and the modulation form rather than the specific realization, the results apply broadly to time-modulated coupled resonator arrays across systems ranging from photonic metamaterials to optical and microwave platforms. 

\section{Theoretical analysis} \label{sec:model}

\subsection{Parametric amplification in a single oscillator} \label{subsec:single}

We first review the dominant parametric resonance of a lone oscillator degree of freedom $x$ with natural frequency $\omega$, upon which we impose a sinusoidal stiffness modulation with relative strength $\delta$ and frequency $\Omega$. The oscillator dynamics are governed by the well-studied Mathieu equation~\cite{LandauLifshitzMechanics,Kovacic2018}
\begin{equation}
  \label{eq:Mathieu_equation}
  \frac{d^2x}{dt^2}+\omega^2[1+\delta \cos{(\Omega t)}] x = 0.
\end{equation}
Floquet theory dictates that \eqnref{eq:Mathieu_equation} has two independent solutions (indexed by $j \in \{1,2\}$) of the form
$$x_j(t) = e^{-i \mu_j t} y_j(t),$$
where $y_j(t)$ are periodic functions of time with period $T = 2\pi/\Omega$, and the $\mu_j$ are termed Floquet frequencies as they dictate the long-time behavior of the solution at multiples of the modulation period: $x_j(t=nT) = e^{-i \mu_j t}x_j(0)$, similar to normal modes for a passive system.
Unlike passive normal mode frequencies, however, the Floquet frequencies can be complex for some values of $\delta$ and $\Omega$, with the real part defining an oscillation frequency modulo $\Omega$, and the imaginary part quantifying the exponential decay or growth of the mode.

Regions of complex $\mu$ in the $(\delta,\Omega)$ phase plane signify instability regions, which can be calculated using a variety of methods~\cite{Kovacic2018}; as $\delta$ approaches zero, these regions narrow to the points $\Omega = 2\omega/m$, $m \in \{1,2,\ldots\}$ which indicate the parametric resonance frequencies.
We will focus on the dominant resonance in the vicinity of $m=1$ or $\Omega = 2\omega$.
At this resonance, the parametric modulation couples the two branches of the harmonic oscillation at frequencies $\pm \omega$ (distinguished by the phase of the velocity relative to the displacement), which is termed \emph{degenerate} amplification.
The internal symmetries of the system guarantee that, at this resonance, the Floquet frequencies of the pair of coupled modes have the form~\cite{Melkani2024}
$$\mu = \frac{\Omega}{2} \pm i s,$$
i.e. the oscillatory part of the Floquet mode has frequency exactly half that of the modulation, and $s$ (which can be chosen to be positive) is the \emph{growth rate} for the exponentially growing mode.
The partner mode amplitude decays with the same rate $s$.
The growing and decaying modes are distinguished by their oscillation phase relative to the modulation.

While the dependence of the growth rate $s$ on the system parameters has no closed form in general, its form at small modulation strengths can be derived using perturbative techniques and has the form~\cite{LandauLifshitzMechanics,Kovacic2018}
\begin{equation} \label{eq:ampfactor}
  s = \frac{1}{2}\left[\left(\frac{\delta \omega}{2}\right)^2-(\Omega - 2 \omega)^2 \right]^{1/2},
\end{equation}
up to terms of higher order in $\delta$.
For amplification to occur, $s$ must be real and positive, which implies that degenerate parametric resonance extends over a range of modulation frequencies satisfying
\begin{equation}
  \label{eq:resonancerange}
  |\Omega - 2\omega| < \frac{\delta}{2}\omega + O(\delta^2).
\end{equation}

To summarize, a stiffness modulation at twice the natural frequency of a harmonic oscillator generates an exponentially growing mode with a growth rate that evaluates to $\delta \omega/4$ at small modulation strengths.
When the modulation frequency is tuned away from $2\omega$, the resonance persists over a finite range of modulation frequencies that depends on the modulation strength (\eqnref{eq:resonancerange}), with a reduced growth rate given by \eqnref{eq:ampfactor}.
We will now adapt these calculations to a complete band of oscillatory modes in a periodic lattice.

\subsection{Periodic array of coupled oscillators}
We consider the lossless linear dynamics of a periodic array of unit cells each with $n$ degrees of freedom, captured by the generic equations of motion
\begin{equation} \label{eq:eom}
    \ddot{\mathbf{u}}_j = -\sum_l\mathbf{K}^{(l-j)}(t)\mathbf{u}_l,
\end{equation}
where $\mathbf{u}$ is the $n$-vector of displacements of each of the masses in unit cell $j$ from their equilibrium position, and $\mathbf{K}^{(l-j)}(t)$ is the $n \times n$ dynamical matrix encoding mass-weighted couplings between unit cell $j$ and unit cell $l$~\footnote{Losses can be incorporated by adding terms that are first-order in time; these generically have the effect of attenuating the strength of the amplification but do not significantly change the resonance conditions~\cite{Kruss2022,Melkani2023a}.}.
Diagonal elements of $\mathbf{K}^{(0)}$ encode the natural oscillation frequencies of the degrees of freedom in the absence of couplings, whereas other matrix elements represent couplings among distinct oscillatory degrees of freedom.
We assume that all interactions are conservative, leading to reciprocal couplings and an overall symmetric dynamical matrix with real entries that satisfy $\mathbf{K}^{(-p)} = (\mathbf{K}^{(p)})^\text{T}$.
We also require that the system is locally stable at every instant, which requires the stiffness matrix to be positive definite at all times.
While we will illustrate our results with a calibrated discrete model of coupled mechanical resonators~\cite{Karki2021}, the framework also applies to continuum systems whose dynamics can be put in the form of \eqnref{eq:eom} through a discretization procedure such as finite-element analysis.

\subsubsection{Floquet-Bloch dispersion relations}
To generate dispersion relations for excitations of the periodic lattice, we first Fourier transform in space, writing
\begin{equation}
	 \mathbf{u}_{j}(t) \propto \sum_{q} e^{i q a j}\,\mathbf{u}_{q}(t),
  	\label{eq:Bloch_transform}
\end{equation}
where $a$ is the periodicity of the lattice in space and $q$ is the wavevector. We then arrive at the equation of motion for each $q$:
\begin{equation}
    \label{eq:uq_eom}
    \frac{d^2\mathbf{u}_q}{dt^2} + \mathbf{K}(q,t)\mathbf{u}_q(t)=0,
\end{equation}
where 
\begin{equation}
    \mathbf{K}(q,t) = \sum_{l}\mathbf{K}^{(l-j)}(t)e^{iqa(l-j)}.
    \label{eq:Fourier_transformed_matrices}
  \end{equation}
The symmetry requirement for the real-space dynamical matrix ensures that $\mathbf{K}(q,t)$ is Hermitian.
When the modulation is periodic in time, $\mathbf{K}(q,t+T) = \mathbf{K}(q,t)$, the Floquet spectrum of $\mathbf{K}(q,t)$ at each $q$ completely describes the  behavior of the system at time scales longer than $T$, and the $2n$ Floquet frequencies $\mu_j(q)$ generate the Floquet-Bloch band structure~\cite{Gomez-Leon2013,Holthaus2015,Rudner2020a} which determines wave propagation at long times in the dynamic structure~\cite{Salerno2016,Vila2017} (see \appref{app:floquetbloch} for details).
In particular, modes with complex Floquet frequencies signify parametrically amplified collective excitations of the array~\cite{Trainiti2019,Kruss2022}.

While the Floquet-Bloch approach enables the numerical evaluation of dispersion relations for any dynamical matrix that is periodic in space and time, it is less effective as a tool to design full-band resonances.
Floquet spectrum calculations involve integrating the equations of motion for all $n$ degrees of freedom in a unit cell, which obscures the physical relationships, outlined in \figref{fig:intro}, that allow for targeted parametric amplification of a particular band.
We now develop a perturbative framework for complete amplification of isolated bands that allows us to generalize the analytical results of \secref{subsec:single} to oscillator arrays.

\subsubsection{Projection onto static bands}
We now restrict ourselves to modulations of the form
\begin{equation}
  \label{eq:timedepK}
  \mathbf{K}(q,t) = \staticK(q) + \delta \cos (\Omega t) \modK(q);
\end{equation}
\emph{i.e.}, we start with a static system with dynamical matrix $\staticK$ and introduce sinusoidal modulations of couplings at a frequency $\Omega$ with modulation strength set by an overall parameter $\delta$.
To start, we assume that the modulation has the same minimal unit as the static system; we will relax this assumption later on.
By specifying a modulation matrix $\modK \neq \staticK$, we allow the different entries in the static dynamical matrix to be modulated by different amounts.
For example, in a system such as in \subfigref{fig:intro}{a} where individual resonators are joined by coupling springs, external fields could modulate the resonance frequencies (vertical springs) differently from the couplings among resonators (horizontal springs).
However, we require that the modulations are small relative to the static couplings, a condition which is imposed by the restrictions $|(\modK)_{ij}| \leq |(\staticK)_{ij}|$ and $0 \leq \delta \leq 1$.
(The sign of $\delta$ is chosen to be positive without loss of generality.) 

The band structure of the static system is obtained by diagonalizing the Hermitian matrix $\staticK$ to provide a set of $n$ orthonormal eigenvectors $\basisvec_\beta$---the Bloch eigenmodes of the system---that satisfy
\begin{equation}
    \staticK(q)\basisvec_\beta(q)=\omega_\beta^2(q)\basisvec_\beta(q),
\end{equation}
with each mode associated with two branches $\pm\omega_\beta(q)$ of the static band structure (as before, the sign distinguishes the phase of the velocity being ahead of or behind the displacement for each mode).
Our goal is to single out a band that is gapped from other bands, and to derive modulation conditions that generate a parametric resonance across the entire band.
To do so, we adapt a standard procedure for multidegree-of-freedom parametric systems---projection onto the normal modes of the static system~\cite{nayfehmook}---to the Bloch eigenmodes which serve as normal modes for the periodic array~\cite{Lellouch2017}.
Specifically, we project the Fourier-space displacements of the dynamic system onto the complete Bloch basis of the static system~\cite{Lellouch2017} via
\begin{equation}
    \mathbf{u}_q(t) = \sum_\beta{x_{\beta q}(t)\basisvec_\beta (q)},
\end{equation}
where the $x_{\beta q}(t)$ are normal mode coordinates.  
The equation of motion
$$\frac{d^2\mathbf{u}_q}{dt^2} + [\staticK(q) + \delta\cos{(\Omega t)}\modK(q)]\mathbf{u}_q(t)=0,$$
when represented using the normal mode coordinates, takes the form
\begin{equation}
  \label{eq:expanded_in_static_basis_eom}
  \begin{split}
    &\sum_\beta{\frac{d^2x_{\beta q}}{dt^2}\basisvec_\beta(q)} + \sum_\beta\omega_\beta^2(q)x_{\beta q}\basisvec_\beta(q)\\ +\, &\delta\cos{(\Omega t)}\sum_\beta\modK(q) x_{\beta q}\basisvec_\beta(q) = 0.
  \end{split}
  \end{equation}
Projecting this equation onto another basis vector, $\basisvec_\alpha(q)$, we obtain
\begin{equation}
    \label{eq:projected_eom_static_basis}
    \frac{d^2x_{\alpha q}}{dt^2}+\omega^2_\alpha x_{\alpha q}+\delta\cos{(\Omega t)}\sum_\beta V_{\alpha\beta}(q)x_{\beta q}=0,
\end{equation}
where $V_{\alpha\beta}(q)=\basisvec_\alpha(q)^\dagger\modK(q) \basisvec_\beta(q)$, and $\dagger$ denotes conjugate transpose.

\Equationref{eq:projected_eom_static_basis} is a key result: it recasts the dynamics of the time-modulated system using independent sets of coupled Mathieu oscillator variables $x_{\alpha q}$, $\alpha \in {1,2,\ldots,n}$ at each wavevector $q$.
The resonance parameters $V_{\alpha\beta}$ capture the strength of the energy exchange between the static modes from bands $\alpha$ and $\beta$ due to the parametric modulation.
Once these parameters are computed, the wealth of techniques developed to analyze coupled Mathieu equations~\cite{nayfehmook} can be applied to provide physical insights into the generation of full-band parametric amplification.
In particular, primary parametric resonances occur among pairs of modes $(a,b)$  whose frequencies satisfy $\Omega \approx \omega_a \pm \omega_b$, and the complex Floquet frequencies of the resonant modes depend only on $V_{ab}$ to linear order in $\delta$.

\subsection{Uniform modulation: amplified flat bands} \label{sec:uniform}
\begin{figure}
    \includegraphics[width=\columnwidth]{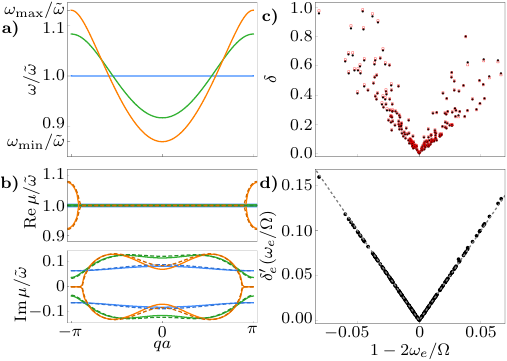}
    \caption{
      Flat amplified bands from uniform modulation.
        \subfiglabel{a} Static dispersion relation for three chains with stiffness parameters chosen to generate varying bandwidths. Frequencies are the square roots of the lowest eigenvalues of $\mathbf{K}_0(q)$, the form of which is given in \appref{app:discrete_model}.
        Only the positive branch of the lowest-frequency band is shown.  Blue: Flat band with $k_1=0.5502$, $k_2=7.3724$. Green: $k_1=0.88032$, $k_2=2.2117$. Orange: $k_1=1.04538$, $k_2=1.47448$. Other parameters are  $\tau=0.5217$, $\kappa=0.0575$ for all three systems.
        \subfiglabel{b} Complex Floquet-Bloch bands for the systems in \subfiglabel{a} upon applying a spatially uniform modulation with modulation parameters  $\delta = 0.7$ and $\Omega = 2\wmid$, where $\wmid = (\wmax+\wmin)/2$.
        Solid curves show Floquet frequencies numerically evaluated as described in \appref{app:floquetbloch}; dashed curves show the perturbative calculation, \eqnref{eq:flat_perturb}. 
        \subfiglabel{c} Minimal modulation strength needed for full-band amplification in 200 distinct systems with static parameters sampled randomly from the ranges $k_1\in(0.385,0.786)$, $k_2\in(5.16,10.53)$, $\tau\in(0.365,0.745)$, $\kappa\in(0.040,0.082)$ to generate a range of bandwidths. Black points: numerical search for threshold $\delta$ (\appref{app:numerical_delta_search}); Red squares: prediction from \eqnref{eq:uniform_threshold}.
        \subfiglabel{d} Numerical search values from panel (c), rescaled according to \eqnref{eq:uniform_threshold} to produce a collapse to the linear prediction $\delta'_e(\omegae/\Omega) = 2|1 - 2\omegae/\Omega|$.
        }
    \label{fig:uniform_modulation}
\end{figure}

For the strategy illustrated in \subfigref{fig:intro}{c}, we aim to tune the modulation frequency $\Omega$ to be near the frequency-doubled resonance, $\Omega \approx 2\omega_1(q)$, for an entire band (indexed as $\alpha = 1$ for convenience).
  To linear order in $\delta$, the equation of motion for the mode amplitude $x_{1q}$ at each wavevector is dominated by the exchange between the two branches $\pm \omega_1(q)$~\cite{nayfehmook}, and \eqnref{eq:projected_eom_static_basis} reduces to
\begin{equation}
    \label{eq:single_mode_Mathieu_equation}
    \frac{d^2x_{1q}}{dt^2}+\omega_1^2(q)\left[1 +  \delta\frac{V_{11}(q)}{\omega_1^2(q)}\cos{(\Omega t)}\right]x_{1q}=0,
\end{equation}
which has the form of the single-mode Mathieu equation (\eqnref{eq:Mathieu_equation}) at each wavevector, but with  $\delta$ replaced by an attenuated $q$-dependent effective modulation strength 
\begin{equation}
  \label{eq:delta_prime_definition}
  \delta'(q) \equiv \delta\frac{V_{11}(q)}{\omega_1^2(q)}.
\end{equation}
Note that if the modulation matrix is identical to the static dynamical matrix,  $V_{11}(q) = \omega_1^2(q)$ and the equation exactly matches the Mathieu form.
Therefore, the attenuation arises due to the mismatch between the normal mode directions of the modulation matrix and those of the static matrix.

With the modified modulation parameter, we can directly apply the perturbative results for the Mathieu equation (\secref{subsec:single}) to \eqnref{eq:single_mode_Mathieu_equation}.
The static pair $\pm \omega_1(q)$ is mapped to a pair of Floquet-Bloch bands with Floquet frequencies
\begin{equation} \label{eq:flat_perturb}
  \mu(q) = \frac{\Omega}{2} \pm \frac{1}{2}\sqrt{\left[\Omega-2\omega_1(q)\right]^2-\left(\frac{\delta'(q)\omega_1(q)}{2}\right)^2} + O(\delta^2),
\end{equation}
with parametric resonance indicated by complex-valued $\mu$.
Thus, the condition for parametric resonance is that the detuning of the static band from half the modulation frequency is smaller than the threshold
\begin{equation} 
  |\Omega - 2\omega_1(q)| < \frac{1}{2}\delta'(q)\omega_1(q).
  \label{eq:resonance_condition_uniform_modulation}
\end{equation}

To amplify a band completely, the condition \eqnref{eq:resonance_condition_uniform_modulation} must be met across the Brillouin zone.
This limits the bandwidth that can be brought into resonance at a given modulation strength, as we show in \subfigref{fig:uniform_modulation}{a--b}.
We generated static bands with varying bandwidths using the calibrated discrete model of plate resonators from Ref.~\onlinecite{Karki2021} (\appref{app:discrete_model}), as quantified by the fractional deviation from the central frequency $\wmid = (\wmax+\wmin)/2$ (\subfigref{fig:uniform_modulation}{a}).
A uniform tension modulation on the plate modifies the resonance frequencies across the array, which we recreate in the discrete model by modulating only the grounding springs (the vertical springs in \subfigref{fig:intro}{a}).
To minimize the detuning across the band, we applied a modulation with $\Omega = 2\wmid$ and the same strength $\delta = 0.7$ across all three systems.
The Floquet-Bloch band structures show complete resonance (nonzero $\text{Im}(\mu)$) for the two bands with lowest bandwidth, while the orange band escapes amplification at the band edges where the detuning from the resonance condition is highest, consistent with the expectation from the Mathieu equation.

The evaluation of the Floquet frequencies illuminates several aspects of the full-band resonance.
First, we confirm the non-perturbative result that the real part of the Floquet band is exactly $\Omega/2$ wherever the imaginary part is nonzero.
Second, although the modulation strength is close to one, we find that the perturbative expression, \eqnref{eq:flat_perturb}, agrees well with the numerically evaluated Floquet frequencies (compare dashed to solid lines in \subfigref{fig:uniform_modulation}{b}).
This shows the impact  of the resonance parameter $V_{11}$, which attenuates the  modulation strength to an effective value $\delta V_{11} /\omega_1^2 \lesssim 0.3$,
where linear perturbation theory is more accurate.
The modulation is generally attenuated since
imposing local stability on the instantaneous dynamical matrices 
ensures $\staticK \pm \modK$ is positive semidefinite, which forces $|V_{11}(q)| \leq \omega_1^2(q)$ for every band and wavevector, with equality only in the case $\modK = \staticK$.
Finally, although the frequency detuning is similar at the band center and the band edges, the resonance parameter attenuates the modulation to a greater degree near $q = \pi/a$ for the range of parameters used here.
Consequently, the imaginary part of the Floquet frequency is lowest at the band edges.

We can use these insights to predict the minimum modulation strength needed to amplify a band across the entire Brillouin zone.
For a given set of system parameters, the threshold modulation strength is that for which \eqnref{eq:resonance_condition_uniform_modulation} is exactly satisfied at $q = \pi/a$. Defining $\omegae \equiv \omega_1(\pi/a)$ and $\delta'_e \equiv \delta'(\pi/a)$ as the frequency and modulation values at the band edge, normalizing by $\Omega$ and rearranging, we find that the threshold modulation strength must satisfy
\begin{equation} \label{eq:uniform_threshold}
  1-\frac{2\omegae}{\Omega}=\pm \frac{\delta'_e\omegae}{2\Omega}.
\end{equation}
We tested this condition using numerical band structure calculations for 200 different realizations of the prototype spring-mass chain, with static parameters sampled randomly from ranges that generate a range of dispersion relations.
Results are reported in \subfigref{fig:uniform_modulation}{c}.      
The black points are from a numerical search (\appref{app:numerical_delta_search}) for the minimum $\delta$ that produces full-band amplification, and the red squares are from the linear prediction obtained by rearranging \eqnref{eq:uniform_threshold} to solve for $\delta$. 
The agreement is generally good, with small deviations at larger bandwidths due to higher-order effects.

The agreement is more apparent in \subfigref{fig:uniform_modulation}{d}, where we test \eqnref{eq:uniform_threshold} directly by plotting the threshold modulation strengths obtained from the numerical search, appropriately rescaled, against the detuning from resonance at the band edge. 
The dashed lines show the prediction from \eqnref{eq:uniform_threshold}, and the values of $\delta'$ collapse onto this line.
The collapse demonstrates that  the typical resonance structure of the Mathieu equation is recovered for Floquet-Bloch bands, as long as the effective modulation strength is used;
the resonance parameter $V_{11}$ fully captures the attenuation of the modulation strength due to the mismatch between the modulation matrix $\modK$ and the normal mode directions of the static matrix $\staticK$.
Deviations at large $\delta'$ show the limits of validity of the linear perturbation theory result, which could be improved by including higher-order terms if necessary.

\subsection{Standing-wave modulation: cosine-like real dispersion relation} \label{sec:standingwave}

\begin{figure}
    \includegraphics[width=\columnwidth]{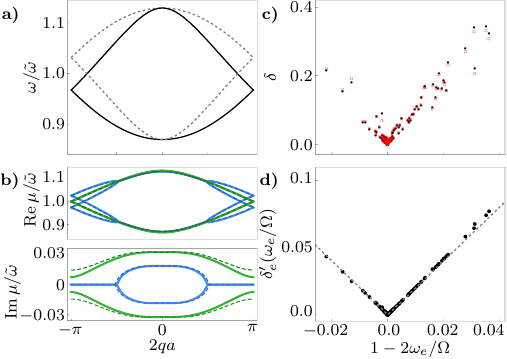}
    \caption{
      Fully amplified band with finite bandwidth using a standing-wave modulation.
      \subfiglabel{a} Folded static dispersion relation (solid line) along with Floquet replica (dashed line). Static parameters are borrowed from the orange system in \subfigref{fig:uniform_modulation}{a}.
      \subfiglabel{b} Real and imaginary components of Floquet frequency bands for standing-wave modulations of the grounding springs (\appref{app:discrete_model}) at two different $\delta$ values: $\delta=0.15$ (blue) and $\delta=0.26$ (green). Solid lines are obtained from the numerical Floquet analysis, while the dashed lines are the perturbative result, \eqnref{eq:linear_two_mode_prediction_of_mu}.
      \subfiglabel{c} Minimal modulation strength $\delta$ needed for full-band amplification in 200 distinct systems with static parameters sampled randomly from the ranges $k_1\in(0.11004, 2.751)$, $k_2\in(1.47448, 36.862)$, $\tau\in(0.104358, 2.60895)$, $\kappa\in(0.0115, 0.2875)$ to generate a range of bandwidths. Red squares are predicted analytical values and black dots are from a numerical search described in \appref{app:numerical_delta_search}.
      \subfiglabel{d} Effective modulation strength vs. edge detuning. Black dots are numerical values of $\delta'_e(\omegae/\Omega)$ and dashed line is the prediction from \eqnref{eq:resonance_condition_band_edge}.
    }	
    \label{fig:two_unit_cells}
\end{figure}

The degenerate parametric amplification and flat amplified bands generated by a uniform modulation ($\staticK$ and $\modK$ sharing the same spatial period $a$) can be avoided by introducing modulations with a different spatial period.
Specifically, if $\modK$ has a period $2a$, with alternating phases between cells of spacing $a$, it couples static modes of $\staticK(q)$ with modes at $\staticK(q+\pi/a) = \staticK(q-\pi/a)$ as shown schematically in \subfigref{fig:intro}{d}~\cite{Cassedy1963,Vila2017,Galiffi2019}.
In the $\delta \to 0$ limit, the primary resonance occurs for static modes that satisfy the resonance condition
\begin{equation}
  \label{eq:sumresonance}
  \omega(q) + \omega(q+\pi/a) \approx \Omega
\end{equation}
across the entire band.
This form departs from that of the uniform resonance in two ways.
First, the resonance is no longer degenerate: the modulation mixes modes at different frequencies, and the real part of the resulting Floquet frequency is no longer required to be a constant across the band at resonance.
Secondly, if a static band has a shifted cosine dispersion relation of the form $\omega(q) = A + B \cos(qa)$, \eqnref{eq:sumresonance} can be satisfied across the entire band upon choosing $\Omega = 2A$; the standing-wave modulation would  completely amplify such a band at arbitrarily small modulation strength.

However, general couplings among resonators with finite frequencies do not generate exact shifted cosines.
Instead, $\cos(qa)$ is often the dominant Fourier component with additional components of the form $\cos(qna)$, $n \in \mathbb{Z}$ also allowed by the periodicity and $q \to -q$ symmetry of the static band as exemplified by the static dispersion relation of the discrete model used here (\appref{app:discrete_model}) and of other waveguide arrays~\cite{Karki2021,Yariv1999,Hutter2011}.
In these cases, the band is detuned from perfect resonance at some wavevectors even if $\Omega$ is chosen to maximize the overlap at others, as illustrated by the gap between the solid and dashed curves in \subfigref{fig:intro}{d}.
To overcome this detuning and generate amplification across the entire band, the  modulation strength again needs to be above a threshold value that increases with the level of detuning.

For exact Floquet frequency calculations, we must identify the periodic unit of the time-dependent dynamical matrix, which is a supercell composed of two unit cells of the static system with lattice period $2a$.
The Brillouin zone for the Floquet-Bloch band structure is the interval $-\pi/(2a) < q \leq \pi/(2a)$.
The modulation matrix  $\modK(q)$ is a $2n \times 2n$ matrix with
terms that are $\pi$ out of phase with each other in the two unit cells comprising the supercell.
The static dynamical matrix $\staticK(q)$ is also a  $2n \times 2n$ matrix, but it simply duplicates the static couplings in the two copies of the unit cell. Full details of the construction of these matrices for the model system are provided in \appref{app:discrete_model}.
As a result, the targeted static band $\omega(q)$ gets folded into two bands in the smaller Brillouin zone (solid line in \subfigref{fig:two_unit_cells}{a}), which we label as $\omega_1(q) = \omega(q)$ and $\omega_2(q) = \omega(q+\pi/a)$.

In the reduced Brillouin zone, the resonance condition \eqnref{eq:sumresonance} is represented by the near-overlap of $\omega_1(q)$ and $\Omega - \omega_2(q)$ in the folded Brillouin zone.
To maximize this overlap across the entire band, we tuned the modulation frequency to satisfy the resonance condition exactly at $q=0$:
$$\Omega = \omega_1(0) + \omega_2(0) = \omega(0) + \omega(\pi/a),$$
which brings the rest of the band nearly into resonance (compare solid and dashed lines in \subfigref{fig:two_unit_cells}{a}).
The resonance is dominated by the pair of coupled equations out of \eqnref{eq:projected_eom_static_basis} that involves $\omega_1$ and $\omega_2$~\cite{nayfehmook}:
\begin{equation}
  \label{eq:two_mode_coupled_Mathieu_equations}
  \begin{split}
    \ddot{x}_1+\omega_1^2 x_1 + \delta\cos{(\Omega t)}V_{12}x_2 &= 0,\\
    \ddot{x}_2+\omega_2^2 x_2 + \delta\cos{(\Omega t)}V_{21}x_1 &= 0;\\
  \end{split}
\end{equation}
where $V_{12}$ and $V_{21}= V_{12}^*$ are computed using the static eigenvectors corresponding to the two bands (see \appref{app:unfolded_v} for details of the computation). 
The pair of Floquet-Bloch bands arising from the coupling between $\omega_1$ and $\omega_2$ can be computed to linear order in the modulation using the method of multiple scales (\appref{app:coupled_mathieu_derivation}) to get 
\begin{equation}
  \label{eq:linear_two_mode_prediction_of_mu}
  \mu = \frac{\omega_1+\Omega-\omega_2}{2} \pm \frac{1}{2}\sqrt{(\Omega-\omega_1-\omega_2)^2 - \frac{\delta^2}{4}\frac{|V_{12}|^2}{\omega_1\omega_2}}
\end{equation}
Thus, the resonance condition for parametric amplification (complex $\mu)$ to occur is
\begin{equation}
  \label{eq:resonance_condition}
  \left|\Omega -[\omega_1(q)+\omega_2(q)]\right| < \frac{\delta}{2}\frac{|V_{12}(q)|}{\sqrt{\omega_1(q)\omega_2(q)}} + O(\delta^2).
\end{equation}

As \subfigref{fig:two_unit_cells}{a} shows, the detuning from the resonance condition is zero at $q=0$ and largest at the band edges $q = \pm \pi/(2a)$ for our modulation scheme.
The resonance is correspondingly strongest at the band center and weaker upon approaching the band edges, as observed in the full Floquet-Bloch spectrum calculations at two modulation levels (solid curves in \subfigref{fig:two_unit_cells}{b}).
At the lower modulation strength, the band is amplified in the vicinity of the band center but not at the band edges; the full band is brought into resonance only with the stronger modulation.
The full spectrum is well-approximated by the perturbative approximation~\eqnref{eq:linear_two_mode_prediction_of_mu} (dashed curves), again confirming the utility of the exact expressions for designing full-band parametric resonances.
Notably, the real Floquet spectrum of the fully amplified band retains the finite bandwidth of the static bands, in contrast to the uniform modulation case which flattened the real spectrum due to the degenerate resonance condition.

As before, we illustrate the utility of the approximate theory by predicting the  modulation strength needed to bring an entire band into resonance.
The limiting resonance again occurs at the band edges, where the folding of the static band $\omega(q)$ produces a degeneracy $\omega_1(\pm\pi/(2a)) = \omega_2(\pm\pi/(2a)) \equiv \omegae$.
[We redefine the band edge frequency compared to the uniform modulation condition, since the band edge now lies at $\pm \pi/(2a)$.]
The minimum $\delta$ needed for full-band resonance satisfies, from \eqnref{eq:resonance_condition},
\begin{equation}
  \label{eq:resonance_condition_band_edge}
  1 - \frac{2\omegae}{\Omega} = \pm \frac{\delta'_e\omegae}{2\Omega},
\end{equation}
where $\delta'_e=\delta|V_{12}(\pi/(2a))|/(\omegae^2)$ is the rescaled modulation strength at the band edge.
Note that the condition is identical in form to the uniform modulation condition \eqnref{eq:uniform_threshold}, but with the frequency and attenuation parameters redefined to reflect the new band edge $q=\pi/(2a)$.

In \subfigref{fig:two_unit_cells}{c}, we compare the threshold $\delta$ obtained numerically from Floquet-Bloch calculations across systems with a wide range of static band parameters (black symbols) against the prediction of \eqnref{eq:resonance_condition_band_edge}.
Plotted against the relative detuning at the band edge $1-2\omegae/\Omega$, the $\delta$ values spread out around two approximately linear branches, and are well-approximated by the perturbative prediction (red symbols), with larger deviations at higher modulation strengths where we expect higher-order terms to become significant.
Upon rescaling the modulation strength by the resonance parameter to define an effective modulation strength $\delta'$, we obtain the collapse onto linear resonance boundaries anticipated by the perturbative prediction (\eqnref{eq:resonance_condition_band_edge}), confirming its validity at low values of $\delta'$.
The resonance parameter, calculated using the spectrum of the static system, allows us to translate insights from the linear analysis of coupled Mathieu equations to complex resonator arrays analogously to the uniform modulation case (\subfigref{fig:uniform_modulation}{c}).

\section{Wave packet manipulation via full-band parametric resonance}

\begin{figure*}[tb]
  \centering
  \includegraphics[width=\textwidth]{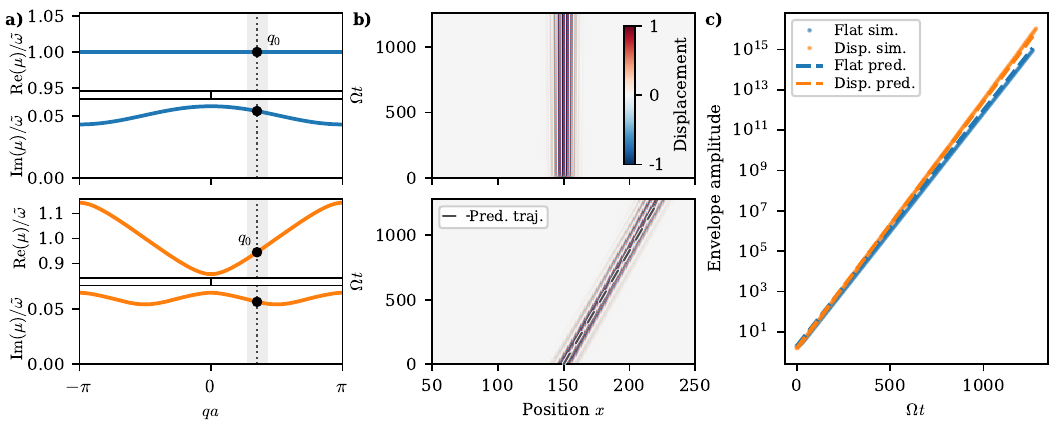}
  \caption{
    Wave packet dynamics in amplified bands.
    \subfiglabel{a} Numerically evaluated Floquet-Bloch band structures. Top: Fully amplified flat band. Static parameters are the same as the blue band in \subfigref{fig:uniform_modulation}{a}, with modulation strength $\delta=0.5$.
    The carrier wavevector $q_0=0.35\pi/a$ used for the packet simulations is highlighted.
    The shaded region indicates the width of the Gaussian packet in wavevector space, $q_0\pm\sigma_q$, which is set by $\sigma_q=0.08\pi/a$.
    Bottom: Fully amplified non-flat band with cosine-like dispersion relation. Static parameters are the same as the green band in \subfigref{fig:two_unit_cells}{b} with modulation strength $\delta = 0.5$, now unfolded to the extended Brillouin zone.
    \subfiglabel{b} Displacement fields of Gaussian wave packets from classical dynamics simulations (see \appref{app:sims} for details) initialized from the Floquet-Bloch modes of the fully amplified bands in \subfiglabel{a}. Displacements at each instant are normalized by the maximum instantaneous displacement. Top: Wave packet from flat band, for which the predicted propagation velocity (slope of $\mathrm{Re}(\mu)$) is zero.
    Bottom: Wave packet initialized from the non-flat  band. Dashed line shows the predicted constant-speed advance with velocity given by the slope of $\mathrm{Re}(\mu)$ at $q_0$.
    \subfiglabel{c} Amplitude of a Gaussian pulse fitted to the wave packets in (b) as a function of time. Dashed lines show the prediction of exponential growth with rate given by the imaginary parts of the Floquet-Bloch bands at $q_0$ [values indicated in \subfiglabel{a}].
    } 
    \label{fig:packets}
\end{figure*}

In \secref{sec:model}, we developed a framework to design full-band parametric resonances in cosine-like bands.
Unlike typical parametric resonances that operate over narrow frequency and wavevector ranges, full-band resonances can be used to amplify and manipulate wave signals with complex spectral characteristics.
We now present examples of such signal processing \emph{via}  dynamical simulations of wave packets propagating through spring-mass chains with system parameters chosen using our theoretical analysis.

In our examples, we simulated the classical dynamics of a chain of 300 unit cells, initialized in a linear superposition of the modes from the static band with Gaussian weights. Specifically, the initial conditions are set up to a normalization factor by
\begin{equation}
  \begin{split}
    \tilde{\mathbf{u}}_j(t) &=
    \sum_{q}
    e^{-(q-q_0)^2/\sigma_q^2}\,\basisvec_1(q)
    e^{i[q(ja-x_0)-\omega_1(q)t]};\\
    \mathbf{u}_j(0) &= \mathrm{Re}\,\tilde{\mathbf{u}}_j(0),
    \qquad 
    \dot{\mathbf{u}}_j(0)
    =
    \mathrm{Re}\,\dot{\tilde{\mathbf{u}}}_j(0),
  \end{split}
\end{equation}
with $q_0$ the carrier wavevector, $\sigma_q$ the width of the Gaussian in wavevector space, and $x_0$ the initial center of the wave packet in real space.
More details of the dynamical simulations are provided in \appref{app:sims}.

\subsection{Amplification of non-propagating and propagating wave packets}

The Floquet-Bloch dispersion relation $\mu(q)$ of the amplified band governs the propagation of wave packets~\cite{Kruss2022}: the slope of the real component sets the propagation speed $v = \partial \text{Re}(\mu) / \partial q$, whereas the imaginary component dictates the exponential growth rate of its constituent modes.
We have shown in \secref{sec:uniform} that under uniform modulation, the degenerate amplified band has a flat real component regardless of the bandwidth of the static band; \emph{i.e.}, its propagation speed is zero (\subfigref{fig:packets}{a}, top panel).
At any carrier frequency $q_0$, a wave packet in a flat amplified band stays in place while growing exponentially in time.
This behavior is confirmed in wave packet simulations, as shown in \subfigref{fig:packets}{b} (top panel) and \subfigref{fig:packets}{c}.
The exponential growth rate of the wave packet amplitude in time matches the value of $\text{Im}(\mu)$ at the carrier frequency. 
We do not see significant distortions over the time interval of our simulation, indicating that the growth rate is sufficiently uniform over the wave vector range of the Gaussian packet to enable high-fidelity amplification~\cite{Kruss2022} (i.e. all wave packet components are amplified at similar rates).

In contrast to the uniform modulation, the non-degenerate standing-wave modulation enabled fully amplified cosine-like bands with a significant bandwidth in the real Floquet frequency (\secref{sec:standingwave}), which allows propagating wave packets to be amplified.
A finite-bandwidth amplified band was demonstrated in \figref{fig:two_unit_cells}, which we unfold into an expanded Brillouin zone in \subfigref{fig:packets}{a}, bottom.
We expect wave packets away from the band center and band edges to propagate with speed given by the slope of $\text{Re}(\mu)$ at the carrier wavevector $q_0$; this expectation is confirmed in simulations of a representative system (\subfigref{fig:packets}{b}, bottom).
The exponential growth of the packet is again predicted by the imaginary part of the Floquet spectrum, as verified in \subfigref{fig:packets}{c}.
Our approach complements the mechanism for amplification of wave packets in gapless bands with linear dispersion relations using traveling-wave modulation in the sonic limit~\cite{Kruss2022}, which generated non-reciprocal amplification.
By using a standing wave rather than a traveling wave, reciprocity is preserved---a wave packet centered at $-q_0$ would move to the left but otherwise show the same amplification behavior as the right-moving wave packet in \figref{fig:packets}.

\subsection{Frequency-preserving temporal wave packet splitting}

\begin{figure}[tb]
  \includegraphics[width=\columnwidth]{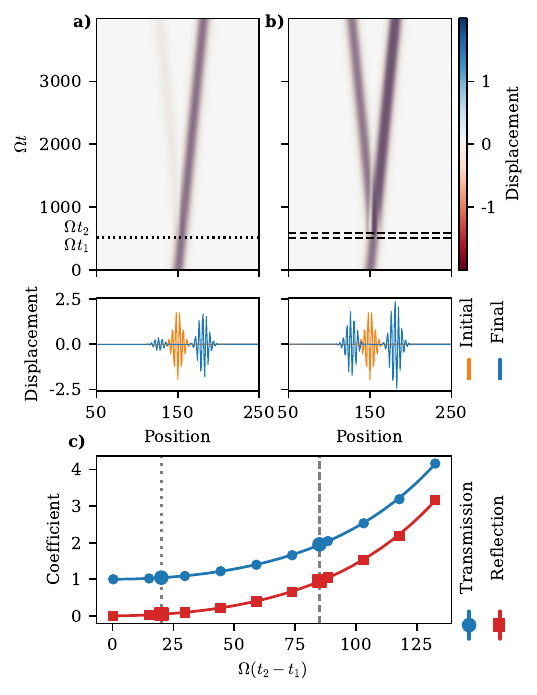}
  \caption{
    Temporal wave packet splitting.
    \subfiglabel{a} Displacement field from simulation of a wave packet with $q_0=0.45\pi/a$ and the static parameters $k_1=0.77028$, $k_2=5.16068$, $\tau=0.5217$, $\kappa=0.0575$. Uniform parametric modulation of the grounding springs with $\Omega=2\wmid$ and $\delta=0.15$ is turned on between $\Omega t_1 = 500$ and $\Omega t_2 = 520$. The lower panel shows the displacements at the initial ($\Omega t=0$) and final ($\Omega t=4000$) time points.
    \subfiglabel{b} Same as (a) but with $\Omega t_2=585$.
    \subfiglabel{c} Transmission and reflection coefficients as a function of the normalized duration of the modulation pulse $\Omega (t_2-t_1)$.
    Symbols are simulation results and solid lines show coefficients predicted from the coupling between forward- and backward-propagating Floquet-Bloch modes generated by the parametric modulation, see \eqnref{eq:transmission_reflection_prediction} and \appref{app:temporal_slab_prediction}. 
    }
    \label{fig:splitting}
\end{figure}

The development of time-modulation as an additional design tool in waveguides has opened up the possibility of wave scattering at \emph{temporal interfaces}: sudden, system-wide changes in parameters that scatter waves in the time domain, akin to regular wave scattering at sharp spatial interfaces~\cite{Xiao2014,Moussa2023,Kim2024,Mendonca2024,Wang2025a,Samak2026}.
While these examples considered sudden changes in otherwise static system parameters, the step change can also involve modifying the characteristics of a periodic parametric modulation, such as its strength or modulation time-period.
In particular, Ref.~\onlinecite{Ye2025} proposed nonreciprocal scattering and amplification of wave packets by turning on a spatiotemporal parametric modulation for a finite time period in an otherwise static elastic beam.
We investigate the effect of such temporal ``slabs'' bound by two sharp temporal interfaces, between which full-band parametric amplification is applied to an otherwise static waveguide. 

Specifically, we constructed a spring-mass chain with static parameters $\staticK$ chosen to have a small but finite bandwidth and thus a nonzero group velocity, similar to the bands in \subfigref{fig:uniform_modulation}{a}.
We then initialized a wave packet using the static modes centered at $q_0 = 0.45 \pi/a$ where the static dispersion relation $\omega(q)$ has a positive slope.
The resulting wave packet propagates towards the right in the static array with constant speed and no amplification or attenuation, as expected (\subfigref{fig:splitting}{a}). 
The wave packet is allowed to propagate up to time $t_1$, when a uniform time-modulation in the form of \eqnref{eq:timedepK} is abruptly turned on with parameters that bring the entire band into resonance.
The time-modulation is maintained at constant $\delta$ and $\Omega$ up to a later time $t_2$, when it is abruptly turned off ($\delta = 0$) and the system is returned to the static parameters.
Since the amplified band has a flat real component, the wave packet does not propagate during the modulation, but instead is modified in-place~\cite{Karki2021}.

We find that  the temporal slab splits the incoming wave packet into two: an outgoing or transmitted packet that propagates in the same direction as the original pulse, and a reflected packet that travels in the opposite direction but at the same speed.
Physically, the uniform modulation couples each mode at $+\omega(q)$ directly to its time-reversed partner at $-\omega(q)$.
Because the resonance is degenerate, both frequency (up to sign) and wavevector are preserved, generating a duplicate wave packet with identical wavevector content but reversed group velocity that propagates as the reflected signal after the modulation is terminated.
Both sets of modes can have energy injected into them by the parametric modulation at resonance, allowing for energy-nonconserving scattering of wave packets~\cite{Ye2025}.
In \subfigref{fig:splitting}{a}, the duration of the modulation $t_2-t_1$ is tuned so that the original wave packet amplitude is preserved despite splitting off a copy, which would be impossible in a passive signal splitter.
In \subfigref{fig:splitting}{b}, the slab duration is increased to generate a reflected pulse with similar amplitude as the transmitted pulse, effectively duplicating the initial signal without losses.

To demonstrate the tunability of the effect, we measured transmission and reflection coefficients (the ratio of total potential energy in the transmitted and reflected packets to that of the initial packet) as a function of the modulation duration (\subfigref{fig:splitting}{c}).
Both coefficients grow with the modulation duration, with the transmission coefficient consistently higher than the reflection coefficient.
This behavior can be understood quantitatively through the parametric pumping mechanism described above, which amplifies both the forward-propagating modes and their time-reversed partners.
For a wave packet narrow in wavevector and centered at $q_0$, the growth rate of its constituent modes is well-approximated by the growth rate of the Floquet-Bloch mode at $q_0$,
\begin{equation}
  s(q_0) = \text{Im}[\mu(q_0)]\equiv s_0>0 .
\end{equation}
A standard calculation using the coupled-mode equations for the modes at $q_0$ (\appref{app:temporal_slab_prediction}) shows that the transmission and reflection coefficients after a modulation duration of $t_{\text{mod}} \equiv t_2-t_1$ are given by
\begin{equation}
  \label{eq:transmission_reflection_prediction}
  \begin{split}
    T_{\text{pred}}(t_{\text{mod}}) & = 1 + \frac{|\lambda|^2}{s_0^2}\sinh^2(s_0 t_{\text{mod}})\\
    R_{\text{pred}}(t_{\text{mod}}) & = \frac{|\lambda|^2}{s_0^2}\sinh^2(s_0 t_{\text{mod}}),
  \end{split}
\end{equation}
with $|\lambda|=\delta V_{11}/(2\Omega)$ the coupling strength between the forward- and backward-propagating modes at $q_0$.
These predictions are plotted as solid lines in \subfigref{fig:splitting}{c}, and show good agreement with the coefficients extracted from the simulated wave packet splitting.

Unsurprisingly, parametric resonance enables coefficients greater than one for both transmission and reflection, illustrating the potential for lossless and amplified wave packet duplication using temporal slabs.
Unlike narrow-band techniques that target specific wave vectors or frequencies for temporal scattering, our approach is designed to manipulate wave packets constructed from any mix of modes in the targeted bands.
For example, a full-band temporal slab can simultaneously duplicate a train of signal pulses, which might be traveling at different group velocities in different sections of the waveguide.

\section{Discussion}

We have demonstrated techniques to fully amplify an isolated band in a one-dimensional system using space-time modulation of material stiffness.
We develop a modeling framework that describes the parametric amplification of the complete band as a system of Mathieu equations with parameters derived solely from the static system and the modulation fields, and is thereby applicable to any parametrically modulated oscillator array for which the static band structure is known.
Our strategies encompass a range of static dispersion relations, from nearly-flat bands to bands with significant spectral width, with tunable reciprocity-preserving dispersion in the real and imaginary parts of the Floquet spectrum.
Unlike typical resonances that target  particular wave frequencies, the full-band parametric resonances generated by our approach can be used to manipulate spectrally complex signals, as we illustrate via numerical demonstrations of wave packet amplification and splitting.

The calculations and simulations presented here describe the linear, lossless, small-signal regime, but any practical amplifier will also be subject to damping.
Under weak, approximately uniform damping, the amplification thresholds are simply shifted by a constant~\cite{Kruss2022}: the effect is analogous to a passive decay rate $\Gamma$ in a Mathieu oscillator (corresponding to an extra term $2\Gamma\dot{x}$ in \eqnref{eq:Mathieu_equation}), which attenuates the growth rate as $s \approx \mathrm{Im}(\mu)-\Gamma$ to linear order~\cite{Kovacic2018}.
Small damping leaves the resonance center essentially unchanged but raises the modulation strength required for net full-band amplification, with the wavevector having the smallest undamped growth rate setting the threshold.
When the damping is significant, amplification may no longer be achievable at practical modulation strengths, but the full-band parametric resonance strategy can still be used to partially compensate for losses by reducing the decay rate of propagating wave signals~\cite{Kruss2022}.

More broadly, parametric resonances provide mechanisms to influence signals even when the modulation strength is not sufficient to overcome damping and all modes are stable---the so-called subthreshold limit.
Although such resonances do not grow spontaneously, they still boost the response of the system to oscillatory forces at the resonance frequency, providing enhanced detection sensitivity to external forcing in single oscillators~\cite{Rugar1991}.
Our design framework provides a route to generating an extensive number of subthreshold resonances in damped parametric oscillator arrays, either at the same resonance frequency (flat amplified band) or with a finite bandwidth (non-flat band).
Such full-band resonances could form the basis for enhanced detection of wave signals that extend along a resonator array, with minimal added noise~\cite{Caves1982,Rugar1991}.

Additionally, our linear analysis is valid only for small amplitudes of the amplified modes.
If the parametric modulation is maintained for long periods, sustained amplitude growth will cause nonlinear effects to become significant and eventually limit the growth of the amplified modes~\cite{nayfehmook}.
Previous works on coupled parametric oscillator networks have found that a balance of parametric amplification and nonlinear effects can lead to a stable saturated response in regimes where the linear theory predicts parametric instabilities~\cite{Lifshitz2003,Kim2023}.
While our general framework for designing full-band amplification remains accurate at the linear level, characterizing the effects of nonlinear saturation and inter-mode coupling on the amplified modes at long times is an important direction for future work.
In particular, the saturated response of a parametrically amplified mode in the nonlinear system is typically bistable between two oscillatory steady states with phases offset by $\pi$~\cite{Lifshitz2003,Lifshitz2008}.
In static nonlinear systems with bistable oscillatory units, noise-driving switching between states gives rise to stochastic resonance phenomena~\cite{Gammaitoni1998} that can enhance the detection of weak periodic signals in arrays of oscillators~\cite{Lindner1995,Gosak2011}.
Our full-band parametric resonances, by providing a macroscopic number of unstable oscillatory states, could support similar noise-enhanced detection mechanisms in the presence of nonlinearities.

We demonstrated our framework using a  discrete model of phonons in membrane resonator arrays in which the background tension can be modulated~\cite{Karki2021,Karki2023}.
Experimentally, such tension modulation has been demonstrated using light and electrostatic fields in individual microresonators~\cite{Mathew2016,Blaikie2019,Miller2020,Carter2023}; extending this tunability to resonator arrays~\cite{Cha2018,Cha2018b} would enable on-device experimental realizations of our approach.
In the near term, larger-scale lattices of mechanical elements with electromagnetically actuated time-modulation~\cite{Wang2018,Kim2023,Kim2024}, provide the most accessible platform for a proof-of-concept realization, since traveling-wave modulation has already been demonstrated in these systems.
Beyond mechanics, a cosine-like dispersion relation is ubiquitous to arrays of oscillator elements such as coupled-resonator optical waveguides for light~\cite{Yariv1999} and Josephson junction arrays for microwaves~\cite{Hutter2011}.
Our design principle summarized in \figref{fig:intro} could be applied to parametric amplification in these platforms, although adjustments will be needed to account for the modified dynamics of the expanded degrees of freedom including nonlinear effects~\cite{OBrien2014,Shi2023}.
The proposed full-band amplification could add to the toolkit for optical and microwave parametric amplification which has versatile applications in  modern communications~\cite{Cerullo2003} and quantum information processing~\cite{Esposito2021}.

Besides expanding our work to other wave manipulation platforms, we also note several promising directions for further investigation.
As mentioned above, the exponentially growing modes will eventually cause nonlinear effects to become significant, which could lead to rich coupled behavior~\cite{Lifshitz2003,Lifshitz2008,Kim2023}.
Understanding the noise characteristics of the fully amplified bands at finite temperature is important for judging amplifier performance, and the noise itself could have interesting spectral characteristics, especially for the uniform modulation which maps to a degenerate parametric amplifier with phase-sensitive noise~\cite{Rugar1991}. 
It would be interesting to consider the effect of moving spatiotemporal interfaces~\cite{Delory2024,Kim2024} generated by phase disruptions in the traveling-wave modulation, which could be enhanced by the presence of a macroscopic number of amplified modes.
Similar mechanisms for full-band parametric amplification also exist for nearly-flat bands in two dimensions~\cite{Karki2023,Samak2024}, and are the subject of ongoing work.

\section*{Data availability}
All graphs are obtained from the numerical evaluation of mathematical operations described in detail in the text, which can be implemented in any scientific computing platform. The specific implementation used to generate the graphs in the manuscript will be made available upon reasonable request.

\begin{acknowledgments}
  \acktext
\end{acknowledgments}

\appendix
\setcounter{figure}{0}
\renewcommand{\thefigure}{A\arabic{figure}}

\section{Floquet-Bloch formulation} \label{app:floquetbloch}

The long-time behavior of systems that are periodically modulated in both space and time is provided by the Floquet-Bloch band structure, which is the Floquet spectrum evaluated for the Fourier-transformed dynamical matrix $\mathbf{K}(q,t)$.
The Floquet-Bloch band structure is widely used in electronic, photonic, and mechanical systems~\cite{Makris2008,Gomez-Leon2013,NassarH.2017,Rudner2020a}.
Here, we recap the method of numerically calculating the Floquet spectrum using direct time-integration of the equations of motion, from Ref.~\onlinecite{Kruss2022}.
These numerically-obtained Floquet frequencies, which are ``exact'' up to the numerical precision of the time-integration, are compared against our analytical results in the main text.
Spectral approaches to compute the Floquet spectrum, which write the equations of motion in frequency space and truncate to a finite number of frequency-Fourier components, could also be used to generate the Floquet frequency bands to the necessary precision and would provide equivalent results.

In its most general formulation, Floquet theory applies to first-order systems of coupled differential equations with periodic coefficients~\cite{YakubovichStarzhinskii}.
Any high-order linear system of ordinary differential equations can be reduced to a first-order system through the introduction of auxiliary variables that represent intermediate derivatives.
These variables are necessary to capture the full dimensionality of the solution space.
To apply Floquet theory to the second-order equations of motion for parametric oscillators (\eqnref{eq:uq_eom}), we define the phase space vector $\mathbf{x} = (\mathbf{u}_q(t), \mathbf{\dot{u}}_q(t))$, so that the equations of motion  can be rewritten as the first-order system~\cite{YakubovichStarzhinskii}:
\begin{equation}
  \label{eq:floquet_de}
  \frac{d\mathbf{x}}{dt} = \textbf{G}(t)\mathbf{x}
\end{equation}
with
\begin{equation}
  \textbf{G}(t) = \begin{pmatrix}
  0 & \textbf{I}\\
  -\textbf{K}(q,t) & 0
  \end{pmatrix},
\end{equation}
where $\textbf{G}(t)$ is periodic with period $T=2\pi/\Omega$ and $\textbf{I}$ is the identity matrix.
Thus, a coupled oscillator system with $n$ degrees of freedom in each unit cell is described by a spanning set of $2n$-dimensional Floquet eigenvectors at each $q$.

Let $\textbf{M}(T)$ be a matrix that evolves the system through one modulation period, i.e. $\mathbf{x}(t+T)=\textbf{M}(T)\mathbf{x}(t)$. Its eigenvalues $\rho$ are the Floquet multipliers, which we parametrize as $\rho=e^{-i\mu T}$, where $\mu$ is the complex Floquet frequency. The corresponding eigenvectors are the Floquet modes, which satisfy $\mathbf{x}(t+T)=\rho\,\mathbf{x}(t)$, so that $\mathrm{Re}(\mu)$ gives the oscillation frequency modulo $\Omega$ and $\mathrm{Im}(\mu)$ gives the exponential growth or decay rate. Parametric amplification therefore appears as bands with nonzero $\mathrm{Im}(\mu)$.

In practice, for each sampled quasimomentum $q$ we compute $\textbf{M}(T)$ as the fundamental matrix solution of \eqnref{eq:floquet_de}. Specifically, we integrate the matrix initial-value problem $\dot{\textbf{M}}(t)=\textbf{G}(t)\textbf{M}(t)$ with $\textbf{M}(0)=\textbf{I}$ from $t=0$ to $t=T$, where this identity initial condition is equivalent to evolving each of the $2n$ basis vectors through one modulation period, using Mathematica's adaptive NDSolveValue routine. The resulting matrix $\textbf{M}(T)$ therefore maps any initial $\textbf{x}(0)$ to its value one period later. We then diagonalize $\textbf{M}(T)$ and convert each multiplier $\rho_j$ to a Floquet frequency using $\mu_j=(i/T)\log\rho_j$, where the branch of the log is chosen so that $\mathrm{Re}(\mu_j) \in [0,\Omega)$. Repeating this procedure across $q$ gives the numerical Floquet-Bloch bands shown in the main text.

When the dynamical matrix is conservative at every instant in time, the spectrum of the time-evolution operator has features constrained by an underlying symplectic structure~\cite{YakubovichStarzhinskii}  which translates to specific symmetries in the complex band structures $\mu(q)$~\cite{Kruss2022,Melkani2023a}:
For every mode with complex frequency $\mu$, there must be a mode with complex frequency $\mu^*$, the complex conjugate of $\mu$.
Due to this symmetry, an entirely resonant band will have two modes that are degenerate in real frequency (modulo $\Omega$) at each quasimomentum, one with a positive imaginary component (signifying amplification) and one with a negative imaginary component (attenuation).
In addition, if the modulation does not break left-right symmetry, then $\mu(q)$ must be equal to $\mu(-q)$.
These features are evident in all the numerical examples presented in the main text.

\section{Discrete model of coupled membrane resonators} \label{app:discrete_model}

The discrete model used here to verify our results was developed to describe the vibrations of arrays of identical membrane resonators with out-of-plane deflections governed by a bending stiffness and a background in-plane tension, $T$, that can be tuned locally or globally~\cite{Karki2021,Karki2023}.

 In the static system, the grounded springs alternate between $k_1$, modeling the fundamental mode of each resonator, and $k_2$, modeling the plate deformation at the junction between resonators. Generically, $k_2 \gg k_1$. Adjacent grounded springs are coupled by a tensed spring with coupling parameter $\tau$.
  There is also a torsional spring with stiffness $\kappa$ that models the bending stiffness at the junction and couples next-nearest-neighbor masses so that the stiffness matrix for an infinite chain is
\begin{equation}
  \textbf{K} =
  \begin{pmatrix}
    ... & ... & ... & ... & ... & ... & ... & ... & ... \\
    ... & \kappa & -c &\alpha& -c & \kappa & 0 & 0 & ...\\
    ... & 0 & 0 & -c & d & -c & 0 & 0 & ...\\
    ... & 0 & 0 & \kappa & -c &\alpha& -c & \kappa &...\\
    ... & 0 & 0 & 0 & 0 & -c & d & -c & ...\\
    ... & ... & ... & ... & ... & ... & ... & ... & ...
  \end{pmatrix}
\end{equation}
with 
\begin{equation}
  \begin{split}
   \alpha& = k_1 + 2\tau + 2\kappa \\
    c & = \tau + 2\kappa \\
    d & = k_2 + 2\tau + 4\kappa.\\
  \end{split}
\end{equation}
An adjacent pair of springs of types 1 and 2 (giving rise to parameters $h$ and $d$) forms a unit cell, with each unit cell representing one membrane resonator in the array. To recreate the effect of modulating the tension within individual resonators, we vary $k_1$ and $k_2$ over time, while keeping $\tau$ and $\kappa$ fixed.

For both the uniform modulation and standing-wave modulation, we have nearest-neighbor coupling between unit cells in the static system and on-site modulation within each unit cell, so the equation of motion for the $j$th unit cell takes the form
\begin{equation}
  \label{eq:uj_eom_uniform}
  \begin{split}
  \ddot{\mathbf{u}}_j = -\staticK^{(-1)}\mathbf{u}_{j-1} - \staticK^{(0)}&\mathbf{u}_j - \staticK^{(1)}\mathbf{u}_{j+1}\\
   - \delta\cos{(\Omega t)}\modK&\mathbf{u}_j,
  \end{split}
\end{equation}
and thus the Fourier-transformed matrices are
\begin{equation}
  \label{eq:specific_fourier_K_matrices}
  \begin{split}
    &\staticK(q) = \staticK^{(-1)}e^{-iqa} + \staticK^{(0)} + \staticK^{(1)}e^{iqa},\\
    &\modK(q) = \modK.
  \end{split}
\end{equation}
The specific forms of these matrices for each modulation are given below.
\subsection{Uniform modulation}
For the uniform modulation, the unit cell remains the same as in the static system, so the stiffness matrices are $2\times 2$ matrices given by
\begin{equation}
  \begin{split}
  &\staticK^{(-1)} = \begin{pmatrix}
    \kappa & -c\\
    0 & 0
  \end{pmatrix}, \qquad
  \staticK^{(0)} = \begin{pmatrix}
    \alpha & -c\\
    -c & d
  \end{pmatrix},\\
  &\staticK^{(1)} = \begin{pmatrix}
    \kappa & 0\\
    -c & 0
  \end{pmatrix},\qquad
  \modK = \begin{pmatrix}
    k_1 & 0\\
    0 & k_2
  \end{pmatrix}.
  \end{split}
\end{equation}

\subsection{Standing-wave modulation with two-unit supercell}
In the case of a standing-wave modulation, the unit cell is doubled. \eqnref{eq:uj_eom_uniform} remains the same, but the stiffness matrices are now $4\times 4$ matrices given by
\begin{equation}
  \label{eq:k_matrices_standing_wave}
  \begin{split}
  &\staticK^{(-1)} = \begin{pmatrix}
    0 & 0 & \kappa & -c\\
    0 & 0 & 0 & 0\\
    0 & 0 & 0 & 0\\
    0 & 0 & 0 & 0
  \end{pmatrix},\quad
  \staticK^{(0)} = \begin{pmatrix}
   \alpha& -c & \kappa & 0\\
    -c & d & -c & 0\\
    \kappa & -c &\alpha& -c\\
    0 & 0 & -c & d
  \end{pmatrix},\\
  &\staticK^{(1)} = \begin{pmatrix}
    0 & 0 & 0 & 0\\
    0 & 0 & 0 & 0\\
    \kappa & 0 & 0 & 0\\
    -c & 0 & 0 & 0
  \end{pmatrix}, \quad
  \modK = \begin{pmatrix}
    k_1 & 0 & 0 & 0\\
    0 & k_2 & 0 & 0\\
    0 & 0 & -k_1 & 0\\
    0 & 0 & 0 & -k_2
  \end{pmatrix}.
  \end{split}
\end{equation}
The periodicity of the doubled unit cell is now $2a$, so \eqnref{eq:specific_fourier_K_matrices} becomes
\begin{equation}
  \begin{split}
    &\staticK(q) = \staticK^{(-1)}e^{-iq(2a)} + \staticK^{(0)} + \staticK^{(1)}e^{iq(2a)},\\
    &\modK(q) = \modK.
  \end{split}
\end{equation}

\section{Numerical search for the minimum modulation strength $\delta$ to achieve full-band resonance} \label{app:numerical_delta_search}
Here we describe the numerical method to find the threshold modulation strengths needed to bring an entire band into resonance, plotted as black symbols in \subfigref{fig:uniform_modulation}{c} and \subfigref{fig:two_unit_cells}{c}.
For each sampled set of static parameters, we first compute the targeted static low band using the numerical Floquet-Bloch evaluation described in \appref{app:floquetbloch} on a uniform grid of $100$ points in wavevector $q$ and record its frequency interval $[\wmin,\wmax]$.
The drive frequency is then fixed by the resonance condition used in the corresponding figure: $\Omega = 2\wmid$ for the uniform modulation and $\Omega = \omega(0) + \omega(\pi/a)$ for the standing-wave modulation.
For the discrete resonator model used here, the last wavevector to come into parametric resonance is at the band edge: $q=\pm\pi/a$ for the uniform modulation and the edge of the reduced Brillouin zone $q=\pm\pi/(2a)$ for the standing-wave modulation.
The threshold search is thus performed at this limiting wavevector.

The search is an upward scan in $\delta$ starting from $0$, with a step size of $\Delta\delta = 2\times 10^{-4}$ for the two-unit supercell and $\Delta\delta = 5\times 10^{-4}$ for the uniform modulation.
At each $\delta$, the Floquet frequencies are computed at the limiting wavevector, and the frequencies corresponding to the target band are isolated by selecting the Floquet modes (modulo $\Omega$) within a small frequency window around the static band frequencies.
Finally, a Floquet mode is considered to be amplified when $\mathrm{Im}(\mu) > 10^{-5}$, and the threshold $\delta$ is recorded when an amplified Floquet mode is first observed at the limiting wavevector.

\section{Calculation of resonance parameters for the standing-wave modulation} \label{app:unfolded_v}

To find the resonance condition for the standing-wave modulation, we work in a basis inherited from the unfolded static system rather than rediagonalizing the folded static problem in the doubled-cell basis.
The static matrix in the unfolded basis is $n \times n$ and has periodicity $\staticK(q+ 2\pi/a) = \staticK(q)$. 
Let $Q$ denote the Bloch wavevector in the reduced Brillouin zone of the doubled unit cell, $-\pi/(2a) < Q \leq \pi/(2a)$.
In the folded basis, both $\staticK(Q)$ and $\modK(Q)$ are $2n \times 2n$ matrices periodic in $Q$ with a period of $\pi/a$. 
The two unfolded static states that fold onto the same $Q$ are the Bloch states at $q=Q$ and $q=Q+\pi/a$.
If $\basisvec(q)$ is an eigenvector of the unfolded static matrix $\staticK(q)$, then the corresponding folded basis vectors can be chosen as
\begin{equation}
  \begin{split}
    &\Phi_1(Q)=\frac{1}{\sqrt{2}}
    \begin{pmatrix}
      \basisvec(Q)\\
      e^{iQa}\basisvec(Q)
    \end{pmatrix}\\
    &\Phi_2(Q)=\frac{1}{\sqrt{2}}
    \begin{pmatrix}
      \basisvec(Q+\pi/a)\\
      -\,e^{iQa}\basisvec(Q+\pi/a)
    \end{pmatrix}.
  \end{split}
\end{equation}
In this basis, the folded static problem restricted to the two lowest modes is already diagonal, with eigenfrequencies $\omega(Q)=\omega_1$ and $\omega(Q+\pi/a)=\omega_2$. 

We now show that in this basis, $V_{11} = V_{22} = 0$ (\eqnref{eq:projected_eom_static_basis}), confirming our expectation that the standing-wave modulation should only couple modes displaced by the modulation frequency $\Omega$ \emph{and} the modulation wavevector $\pi/a$, which excludes self-resonance. To do so, we define a translation operator that translates the lattice by the basic lattice spacing $a$,
\begin{equation}
  \mathcal{T}_a = \begin{pmatrix}
    0 & \mathbf{I}_n\\
    e^{2iQa}\mathbf{I}_n & 0
  \end{pmatrix},
\end{equation}
where $\mathbf{I}$ is the $n \times n$ identity matrix, and $n$ is the number of degrees of freedom in each basic unit cell. Note that $\mathcal{T}_a^2 = e^{2iQa}\mathbf{I}$, which is exactly translation by the extended unit cell spacing 2a.

The folded basis vectors are eigenvectors of $\mathcal{T}_a$:
\begin{equation}
  \begin{split}
    &\mathcal{T}_a\Phi_1(Q) = e^{iQa}\Phi_1(Q)\\
    &\mathcal{T}_a\Phi_2(Q) = -e^{iQa}\Phi_2(Q).
  \end{split}
\end{equation}

The condition for a standing-wave modulation is that the modulation is advanced by a phase of $\pi$ upon translation by $a$. Therefore, a way of defining this condition mathematically is to require that the effect of applying the translation operator to the stiffness matrix in \eqnref{eq:timedepK} is to advance the phase inside the cosine by $\pi$:
\begin{equation}
  \begin{split}
    \mathcal{T}_a^\dagger\mathbf{K}(Q,t)\mathcal{T}_a &= \mathbf{K}(Q,t+\pi/\Omega)\\
    &=\staticK(Q) + \delta\cos{(\Omega t + \pi)}\modK(Q)\\
    &=\staticK(Q) - \delta\cos{(\Omega t)}\modK(Q).
  \end{split}
\end{equation}
This implies that $\mathcal{T}_a^\dagger\modK(Q)\mathcal{T}_a=-\modK(Q)$, where $\modK(Q)$ is the modulation matrix in the extended unit cell basis, given by \eqnref{eq:specific_fourier_K_matrices} and \eqnref{eq:k_matrices_standing_wave} for our model.
Similarly, $\mathcal{T}_a\modK(Q)\mathcal{T}_a^\dagger=-\modK(Q)$.
Using this relation, we can show that the diagonal matrix elements of the modulation vanish in the folded basis:
\begin{equation}
  \begin{split}
    V_{\alpha\alpha} = \Phi_\alpha^\dagger\modK(Q)\Phi_\alpha
    &=
    (\mathcal{T}_a\Phi_\alpha)^\dagger
    \mathcal{T}_a\modK(Q)\mathcal{T}_a^\dagger
    (\mathcal{T}_a\Phi_\alpha)\\
    &=
    -\Phi_\alpha^\dagger\modK(Q)\Phi_\alpha\\
    &= -V_{\alpha\alpha},
  \end{split}
\end{equation}
since $\mathcal{T}_a\Phi_\alpha$ differs from $\Phi_\alpha$ only by a phase.
Thus $V_{11}=V_{22}=0$ and the only nonzero coupling terms are $V_{12}$ and $V_{21}$, given by 
\begin{equation}
  \begin{split}
    &V_{12}(Q)=\Phi_1^\dagger(Q)\,\modK(Q)\,\Phi_2(Q)\\
    &V_{21}(Q)=\Phi_2^\dagger(Q)\,\modK(Q)\,\Phi_1(Q).
  \end{split}
\end{equation}
Because $\modK(Q)$ is Hermitian, $V_{12}^* = V_{21}$ so we have $V_{12}V_{21} = |V_{12}|^2$.

\section{Perturbation theory of Floquet frequencies under standing-wave modulation} \label{app:coupled_mathieu_derivation}

Here, we derive the perturbative expressions for the Floquet frequency to linear order in the modulation strength for the nondegenerate resonance due to the standing-wave modulation.
These results have been derived previously (e.g.~\cite{nayfehmook}) but are reproduced here for completeness.

We begin with an ansatz for the solution of the form
\begin{equation}
  \label{eq:ansatz_for_linear_mu_prediction}
  x_1(t) = ae^{-i\mu t}, \quad x_2(t) = be^{i(\Omega-\mu) t},
\end{equation}
since the resonance occurs between mode 1 and the copy of mode 2 displaced by the modulation frequency $\Omega$. Here $a$ and $b$ are complex constants and $\mu$ is the complex Floquet frequency to be determined.
Substituting the ansatz into \eqnref{eq:two_mode_coupled_Mathieu_equations} and ignoring fast oscillating terms (frequency $\Omega+\mu$ and $2\Omega - \mu$), we obtain:
\begin{equation}
  \label{eq:coefficient_equations_for_linear_mu_prediction}
  \begin{split}
    (\omega_1^2-\mu^2)a + \frac{\delta}{2}V_{12}b &= 0\\
    [\omega_2^2-(\Omega-\mu)^2]b + \frac{\delta}{2}V_{21}a &= 0,\\
  \end{split}
\end{equation}
which has a nontrivial solution when
\begin{equation}
  \label{eq:determinant_equation_for_linear_mu_prediction}
  (\omega_1^2-\mu^2)[\omega_2^2-(\Omega-\mu)^2] = \frac{\delta^2}{4}V_{12}V_{21}.
\end{equation}
Defining small parameters $\zeta=\mu-\omega_1$ and $\sigma=\Omega-(\omega_1+\omega_2)$, we find that
\begin{equation}
  \label{eq:small_parameter_expansion_for_linear_mu_prediction}
  \begin{split}
    \omega_1^2-\mu^2 & = -2\omega_1 \zeta + O(\zeta^2)\\
    \omega_2^2-(\Omega-\mu)^2 & = -2\omega_2(\sigma-\zeta) + O(\zeta^2, \sigma^2, \zeta\sigma).
  \end{split}
\end{equation}
With these expressions, \eqnref{eq:determinant_equation_for_linear_mu_prediction} reduces to
\begin{equation}
  \zeta^2 - \sigma \zeta + \frac{\delta^2}{16}\frac{V_{12}V_{21}}{\omega_1\omega_2} + O(\zeta^3, \sigma^3, \zeta^2\sigma, \zeta\sigma^2) = 0.
\end{equation}
Solving this quadratic equation for $\zeta$ and then plugging back in to $\mu = \omega_1 + \zeta$ gives us the linear prediction for the Floquet frequency,
\begin{equation}
  \label{eq:appendix_linear_two_mode_prediction_of_mu}
  \mu = \frac{\omega_1+\Omega-\omega_2}{2} \pm \frac{1}{2}\sqrt{(\Omega-\omega_1-\omega_2)^2 - \frac{\delta^2}{4}\frac{V_{12}V_{21}}{\omega_1\omega_2}},
\end{equation}
and the resonance boundary for parametric amplification (i.e. the threshold at which $\mathrm{Im}(\mu) \neq 0$) is given by
\begin{equation}
  \label{eq:appendix_two_mode_resonance_condition}
  \Omega = \omega_1+\omega_2 \pm \frac{\delta}{2}\left(\frac{V_{12}V_{21}}{\omega_1\omega_2}\right)^{1/2}+O(\delta^2).
\end{equation}

\section{Wave packet simulations} \label{app:sims}
The dynamical wave-packet simulations in \figref{fig:packets} and \figref{fig:splitting} use the same two-degree-of-freedom discrete resonator model described in \appref{app:discrete_model} in a finite system with $N=300$ unit cells and periodic boundary conditions.
The displacement vector is
\begin{equation}
  \mathbf{y}(t) =
  (u_{1,0},u_{2,0},\ldots,u_{1,N-1},u_{2,N-1})^T,
\end{equation}
where $u_{1,\ell}$ and $u_{2,\ell}$ are the displacements of the first and second oscillators in unit cell $\ell$, respectively.
At each time step, the stiffnesses are evaluated as
\begin{equation}
  \begin{split}
    k_{1,\ell}(t) &= k_1\left[1+\delta(t)\cos(g a\ell-\Omega t)\right],\\
    k_{2,\ell}(t) &= k_2\left[1+\delta(t)\cos(g a\ell-\Omega t)\right],
  \end{split}
  \label{eq:app_sim_modulated_springs}
\end{equation}
with $g=0$ for a spatially uniform modulation and $g=\pi/a$ for the two-cell standing-wave modulation.
The acceleration used in the simulations is then
\begin{equation}
\mathbf{a}(t) =
  (\ddot{u}_{1,0},\ddot{u}_{2,0},\ldots,\ddot{u}_{1,N-1},\ddot{u}_{2,N-1})^T,
\end{equation}
with
\begin{equation}
  \begin{split}
    m\ddot{u}_{1,\ell} = -\big[&
    (k_{1,\ell}(t)+2\tau+2\kappa)u_{1,\ell}
    +\kappa(u_{1,\ell-1}+u_{1,\ell+1})\\
    &-(\tau+2\kappa)(u_{2,\ell-1}+u_{2,\ell})\big],
  \end{split}
  \label{eq:app_sim_u1_accel}
\end{equation}
and
\begin{equation}
  \begin{split}
    m\ddot{u}_{2,\ell} =
    &-[
      (k_{2,\ell}(t)+2\tau+4\kappa)u_{2,\ell}\\
      &-(\tau+2\kappa)(u_{1,\ell}+u_{1,\ell+1})
    ].
    \label{eq:app_sim_u2_accel}
  \end{split}
\end{equation}
For periodic boundary conditions, indices are wrapped modulo $N$.
For open boundary conditions, displacements and parameters outside the chain are set to zero.
The simulations shown in \figref{fig:packets} and \figref{fig:splitting} use $N=300$, $a=1$, $m=1$ and periodic boundary conditions.

The equations of motion are integrated using a velocity Verlet scheme \cite{tuckerman2010}.
Let $\Delta t$ be the integration step and $\mathbf{a}_n=\mathbf{a}(\mathbf{y}_n,t_n)$ the acceleration computed from \eqnref{eq:app_sim_u1_accel} and \eqnref{eq:app_sim_u2_accel}.
The update is
\begin{equation}
  \begin{split}
    \mathbf{y}_{n+1} &=
    \mathbf{y}_n+\Delta t\,\mathbf{v}_n
    +\frac{1}{2}\Delta t^2\,\mathbf{a}_n,\\
    \mathbf{v}_{n+1} &=
    \mathbf{v}_n
    +\frac{1}{2}\Delta t\,\left[\mathbf{a}_n+
    \mathbf{a}(\mathbf{y}_{n+1},t_{n+1})\right].
  \end{split}
  \label{eq:app_sim_verlet}
\end{equation}
All manuscript wave-packet figures use $\Delta t=10^{-2}$ and no damping.
Snapshots are recorded every few integration steps for plotting.

For the temporal-slab simulations in \figref{fig:splitting}, the system is initialized in the static lattice and the modulation is applied only during a finite time window.
The schedule is implemented by setting $\delta(t)=0$ outside the pulse and $\delta(t)=0.15$ inside it, with $g=0$ and $\Omega=2\wmid$.
The two space-time panels use pulse durations $\Omega(t_2-t_1)=20$ and $85$, and the coefficient plot is obtained by sweeping the pulse duration from $0$ to $130$.
The transmitted and reflected weights are estimated from the final displacement profile by summing $u_{1,\ell}^2$ and $u_{2,\ell}^2$ over the two halves of the chain and normalizing by the initial displacement weight.

\section{Prediction of temporal-slab transmission and reflection coefficients} \label{app:temporal_slab_prediction}
Here we describe the prediction for the transmission and reflection coefficients in \subfigref{fig:splitting}{c}. 
The following is the standard slow-amplitude description of the primary parametric resonance of the Mathieu equation~\cite{LandauLifshitzMechanics,nayfehmook,Kovacic2018}, equivalently written as the degenerate parametric-amplifier coupled-mode equations~\cite{louisell}.
The prediction treats the incident wave packet as a narrow packet centered at one Bloch wavevector $q_0$, and then applies the standard degenerate parametric-amplifier result to that single carrier mode. 
This gives the mechanism responsible for the observed growth and is a good approximation as long as the growth rate is relatively constant near $q_0$.

For a spatially uniform modulation, the Bloch wavevector $q$ is conserved during the modulation. We therefore focus on the static Bloch mode at the carrier wavevector $q_0$ and consider it as a parametrically driven oscillator,
\begin{equation}
  \label{eq:app_temporal_slab_prediction_parametric_oscillator}
  \ddot{x} + \omega^2[1+\delta'\cos(\Omega t)]x = 0,
\end{equation}
with $\omega = \omega_1(q_0)$, $x=x_{1q_0}$, and $\delta' = \delta V_{11}/\omega^2$. We set $\Omega = 2\omega + \Delta$ to be near resonance. Let
\begin{equation}
  \label{eq:app_temporal_slab_prediction_nu_definition}
  \nu = \Omega/2 = \omega + \Delta/2.
\end{equation}
Then we write an ansatz for the solution:
\begin{equation}
  \label{eq:app_temporal_slab_prediction_ansatz}
  x(t) = a(t)e^{-i\nu t} + b(t)e^{i\nu t},
\end{equation}
where $a(t)$ and $b(t)$ are slowly varying envelopes. Substituting this ansatz to the static part of \eqnref{eq:app_temporal_slab_prediction_parametric_oscillator}, we obtain
\begin{equation}
  \label{eq:app_temporal_slab_prediction_static_part}
  \begin{split}
  \ddot{x} + \omega^2 x = &[(\omega^2-\nu^2)a-2i\nu \dot{a}]e^{-i\nu t}\\
   + &[(\omega^2-\nu^2)b+2i\nu \dot{b}]e^{i\nu t} + \text{terms}\propto\ddot{a},\ddot{b}.
  \end{split}
\end{equation}
The resonant part of the modulation term is
\begin{equation}
  \label{eq:app_temporal_slab_prediction_modulation_part}
  \delta'\omega^2\cos(\Omega t)x = \frac{\delta'\omega^2}{2}(a e^{i\nu t} + b e^{-i\nu t}) + \text{fast terms}.
\end{equation}
Keeping terms to first order in the detuning $\Delta$ and equating coefficients of $e^{-i\nu t}$ and $e^{i\nu t}$ gives us the coupled equations for the envelopes,
\begin{equation}
    \begin{pmatrix}
        \dot{a} \\
        \dot{b}
    \end{pmatrix}
    =
    \mathbf{M}
    \begin{pmatrix}
        a \\
        b
    \end{pmatrix}, \quad
    \mathbf{M} =
    \begin{pmatrix}
        i\Delta/2 & \lambda \\
        \lambda^* & -i\Delta/2
    \end{pmatrix},
    \label{eq:app_temporal_slab_prediction_coupled_equations}
\end{equation}
where $\lambda = -i\delta'\omega^2/(4\nu) = -i\delta V_{11}/(2\Omega)$.
The evolution of $a$ and $b$ after time $t$ is obtained by integrating the coupled equations above, which gives
\begin{equation}
    \label{eq:app_temporal_slab_prediction_coupled_equations_solution}
    \begin{pmatrix}
        a(t) \\
        b(t)
    \end{pmatrix}
    =
    e^{\mathbf{M}t}
    \begin{pmatrix}
        a(0) \\
        b(0)
    \end{pmatrix}
\end{equation}

The matrix exponential can be simplified by writing $\mathbf{M}$ using Pauli matrices:
\begin{equation}
    \mathbf{M}
    =
    \lambda_R\sigma_x-\lambda_I\sigma_y+i\frac{\Delta}{2}\sigma_z
    \equiv \mathbf{d}\cdot\boldsymbol{\sigma},
\end{equation}
where $\mathbf{d}=(\lambda_R,-\lambda_I,i\Delta/2)$ and $\lambda=\lambda_R+i\lambda_I$.
The standard Pauli exponential identity then gives
\begin{equation}
    \label{eq:app_temporal_slab_prediction_exponential_of_M}
    e^{\mathbf{M}t}
    =
    \mathbf{I}\cosh(s_0 t)
    +
    \frac{\mathbf{M}}{s}\sinh(s_0 t),
  \end{equation}
  where $s_0 = \sqrt{\mathbf{d} \cdot \mathbf{d}}= \sqrt{|\lambda|^2-\frac{\Delta^2}{4}}$ is the  growth rate of the resonant mode.

At the beginning of the temporal slab we take the incident packet to contain only the forward, positive-frequency component. The small mismatch between $\nu$ and the unmodulated frequency $\omega$ only appears as a slow phase in the incident envelope, so choosing the start of the slab as $t=0$, this phase is absorbed into $a_{\mathrm{in}}$. So the initial conditions for the envelopes are
\begin{equation}
    a(0)=a_{\mathrm{in}}, \qquad b(0)=0 .
\end{equation}
Plugging in to \eqnref{eq:app_temporal_slab_prediction_coupled_equations_solution} along with \eqnref{eq:app_temporal_slab_prediction_exponential_of_M} gives us the envelopes after modulation time $t$:
\begin{equation}
  \begin{split}
    a(t)
    &=
    a_{\mathrm{in}}
    \left[
    \cosh(s_0 t)
    +
    i\frac{\Delta}{2s_0}\sinh(s_0 t)
    \right],\\
    b(t)
    &=
    a_{\mathrm{in}}
    \frac{\lambda^*}{s_0}\sinh(s_0 t).
  \end{split}
\end{equation}

The transmitted and reflected weights are then given by
\begin{equation}
  \begin{split}
    T_{\text{pred}}(t) & = \frac{|a(t)|^2}{|a_{in}|^2} = 1 + \frac{|\lambda|^2}{s_0^2}\sinh^2(s_0 t)\\
    R_{\text{pred}}(t) & = \frac{|b(t)|^2}{|a_{in}|^2} = \frac{|\lambda|^2}{s_0^2}\sinh^2(s_0 t).
  \end{split}
\end{equation}
At exact resonance ($\Delta=0$), the growth rate is $s_0 = |\lambda|$ and these reduce to
\begin{equation}
  T_{\text{pred}}(t) = \cosh^2(|\lambda| t), \quad R_{\text{pred}}(t) = \sinh^2(|\lambda| t).
\end{equation}
Note that this calculation is precisely related to the Floquet exponent $\mu$ in that $\mathrm{Re}(\mu) = \nu$ and $\mathrm{Im}(\mu) = s_0$. Thus, the growth of the transmitted and reflected weights is directly related to the growth rate of the modes near $q_0$.

\end{document}